\documentclass[aps,prl,twocolumn,notitlepage,superscriptaddress]{revtex4-2}

\usepackage[T1]{fontenc}
\usepackage[utf8]{inputenc}
\usepackage{lmodern}           
\usepackage{microtype}         

\usepackage{amsmath,amssymb,amsfonts,mathtools,bm,mathrsfs}

\usepackage{siunitx}
\DeclareSIUnit{\pg}{pg}
\DeclareSIUnit{\GHz}{GHz}

\usepackage{graphicx}
\graphicspath{{./figures/}{./}}

\usepackage{tikz}
\usetikzlibrary{shapes.geometric,arrows,decorations.markings}

\usepackage[dvipsnames]{xcolor}

\usepackage[normalem]{ulem}    
\usepackage{enumitem}          

\makeatletter
\def\@seccntformat#1{\csname the#1\endcsname.\,\ }     
\makeatother

\usepackage[colorlinks=true,linkcolor=blue,citecolor=magenta,urlcolor=blue]{hyperref}
\allowdisplaybreaks

\begin{document}

\title{Bilayer Cuprate Antiferromagnets Enable Programmable Cavity Optomagnonics}

\author{Tahereh Sadat Parvini}
\email{tahereh.parvini@uni-greifswald.de}
\affiliation{Walther-Meißner-Institut, Bayerische Akademie der Wissenschaften, Walther-Meißner-Str.8, 85748 Garching, Germany}
\affiliation{Munich Center for Quantum Science and Technology (MCQST), Schellingstr.4, 80799, Munich, Germany}

\date{\today}

\begin{abstract}
Hybrid platforms that couple microwave photons to collective spin excitations offer promising routes for coherent information processing, yet conventional magnets face inherent trade-offs among coupling strength, coherence, and tunability. We demonstrate that bilayer cuprate antiferromagnets—exemplified by YBa$_2$Cu$_3$O$_{6+x}$—provide an alternative approach enabled by their unique magnon spectrum. Using a neutron-constrained bilayer spin model, we obtain the complete $\Gamma$-point spectrum and identify an in-plane acoustic $\alpha$ mode that remains gapless and Zeeman-linear, alongside an in-plane optical $\beta$ mode stabilized by weak anisotropy whose frequency can be tuned from GHz to THz ranges. When coupled to a single-mode microwave cavity, these modes create two distinct channels with a magnetically tunable $\alpha$--photon coupling where $\lvert\mathcal{G}_\alpha\rvert \propto B^{-1/2}$ and a nearly field-independent $\beta$--photon coupling. This asymmetric behavior enables continuous, single-parameter control spanning from dispersive to strong coupling regimes. In the dispersive limit, Schrieffer--Wolff analysis reveals cavity-mediated magnon--magnon coupling $g_{\alpha\beta}^{\mathrm{eff}} \propto g_\alpha g_\beta/(\omega_m-\omega_c)$, while near triple resonance ($f_c = f_\alpha = f_\beta$) the normal modes reorganize into bright and dark superpositions governed by a single collective energy scale. The calculated transmission exhibits vacuum-Rabi splittings, dispersive shifts, and Fano-like lineshapes that provide concrete experimental benchmarks and suggest potential for programmable filtering and coherent state transfer across the GHz-THz frequency range if realized experimentally with suitable interfaces.
\end{abstract}

\maketitle
\section{Introduction}

Engineered light-matter interactions enable transformative advances across classical photonics~\cite{wang2022optical, parvini2017new, pontula2025non, badloe2024enabling, winkel2024comparative, parvini2022magnetooptical, parvini2015kolakoski}, neuromorphic computing architectures~\cite{tait2017neuromorphic,li2025photonics, prucnal2017neuromorphic, oberbauer2025magnetic, shashank2025bulk}, and quantum information technologies~\cite{chan2023chip, sekine2024microwave, liu2025integrated, zhang2023review, ying2023nodes, psaltis2002coherent}. Cavity magnonics—where magnons hybridize with microwave and optical photons—enables tunable quantum dynamics~\cite{hirayama2021hybrid, song2025single, ghirri2023ultrastrong, lee2023cavity, rani2025high, liu2016optomagnonics, rameshti2022cavity, osada2018brillouin, liang2023all, haigh2016triple, simic2020coherent, zhang2014strongly, hisatomi2016bidirectional}. Yttrium iron garnet (YIG)~\cite{lownoise, soykal2010strong, soykal2010size, huebl2013high, tabuchi2014hybridizing} established the benchmark with cooperativity exceeding $10^{5}$~\cite{goryachev2014high, bourhill2016ultrahigh, xu2024nonreciprocal}, multi-GHz splittings, multimode hybridization involving photons, magnons, and phonons~\cite{zuo2024cavity, shen2025cavity, tabuchi2016quantum}, and Gilbert damping $8.6\times10^{-5}$~\cite{guo2023strong}, achieving single-magnon detection and quantum-limited sensing. YIG still faces engineering challenges, including substrate-driven trade-offs between mode volume and coherence, increased millikelvin damping from paramagnetic substrates, and a narrower accessible frequency range than antiferromagnets~\cite{HybridLI, soykal2010size}, motivating exploration of alternative magnetic platforms.

Antiferromagnetic (AFM) insulators have transformed cavity magnonics through fundamental advantages over ferrimagnets: negligible stray fields, ultrafast GHz-THz dynamics, and robust frequency tunability~\cite{lachance2019hybrid, parvini2024cavity, parvini2020, frostad2025stability, kaarbo2025strong, falch2025second, shiranzaei2023temperature}. Recent terahertz-cavity experiments with NiO have realized cavity magnon–polaritons in the THz band~\cite{kritzell2024terahertz, moriyama2019intrinsic}, and Hematite ($\alpha$-Fe$_2$O$_3$) has shown room-temperature strong magnon–photon coupling, underscoring that AFMs can operate at high frequencies with low loss~\cite{bialek2021strong, boventer2023antiferromagnetic, yang2022quantum, ghirri2023ultrastrong, bialek2023cavity}. Cuprate AFMs—exemplified by YBa$_2$Cu$_3$O$_{6+x}$ (YBCO) and Bi$_2$Sr$_2$CaCu$_2$O$_{8+x}$ (BSCCO)—are paradigmatic correlated materials in which intertwined spin, charge, and lattice degrees of freedom support high-$T_c$ superconductivity and robust antiferromagnetism~\cite{tranquada1988, singh1992role, ajay1995effect, pratap1999magnetic,jin2025exploring, bounoua2022hidden, reznik2006spin, le2011intense, Tranquada2007, bao2025magnon}. Their spin dynamics are set by a remarkably large in-plane superexchange, $J_{\parallel}\!\sim\!130~\mathrm{meV}$, and quasi-two-dimensional correlations that persist well above $T_N$~\cite{haug2010neutron, shamoto1993neutron}. These extreme energy scales, together with the bilayer structure of YBCO, make cuprates natural candidates for cavity magnonics and optomagnonics. Theory predicts that cuprates embedded in THz cavities can realize both linear and quadratic photon–magnon interactions, including a bilayer-specific bimagnon channel~\cite{curtis2022cavity, Ghirri}. These advances position cuprate AFMs as a promising cavity-QED platform.

In this work, we develop a microwave cavity–optomagnonic platform based on bilayer cuprate AFMs, with a focus on YBCO. Starting from a neutron-constrained bilayer spin Hamiltonian (Sec. \ref{model}), we obtain the full magnon spectrum and, at the Brillouin-zone center $\Gamma$, identify an in-plane acoustic Goldstone mode $\alpha$ (gapless at $H{=}0$, Zeeman-linear) and an in-plane optical mode $\beta$ whose gap is set by the easy-plane anisotropy $\alpha_D$ and is therefore field-stiff. Reducing $\alpha_D$ lowers $f_\beta$ and shifts the $\alpha$–$\beta$ crossing to lower frequencies. Coupling a single microwave cavity mode (Sec. \ref{Optomagnonic}) yields a beam-splitter interaction with two channels: a field-tunable $\alpha$–photon coupling scaling as $|\mathcal G_\alpha|\propto\sqrt{f_c/H}$ and a nearly field-independent $\beta$–photon coupling. In Sec. \ref{Spectroscopy} we formulate an input–output framework for the transmission $S_{21}(\omega)$ in three configurations: (i) probe locked to $\alpha$, realizing purely dissipative loading and magnetic/cavity-frequency control of linewidth and transmission; (ii) probe locked to the cavity, operating as a field-programmable notch or transparency window; and (iii) frequency sweeps at fixed $f_c$, revealing vacuum–Rabi doublets at resonance and Fano-like lineshapes away from it, with the far-detuned $\beta$ branch contributing only a weak dispersive background.

\section{Bilayer Cuprate AFM Hamiltonian}
\label{model}

\begin{figure*}[hpt!]
    \centering
    \includegraphics[width=0.34\textwidth]{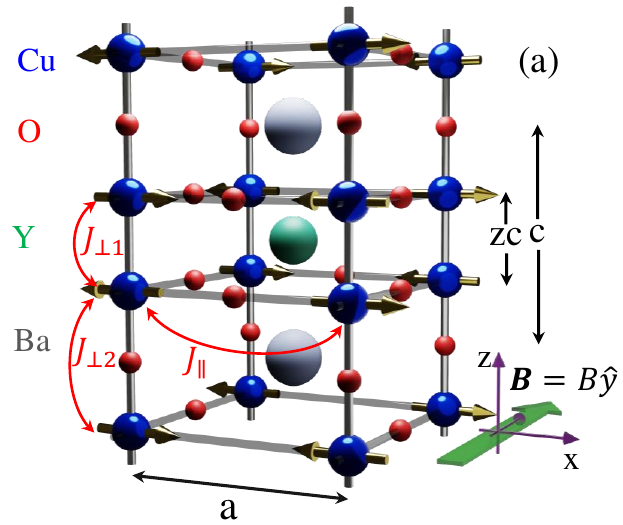}
    \includegraphics[width=0.39\textwidth]{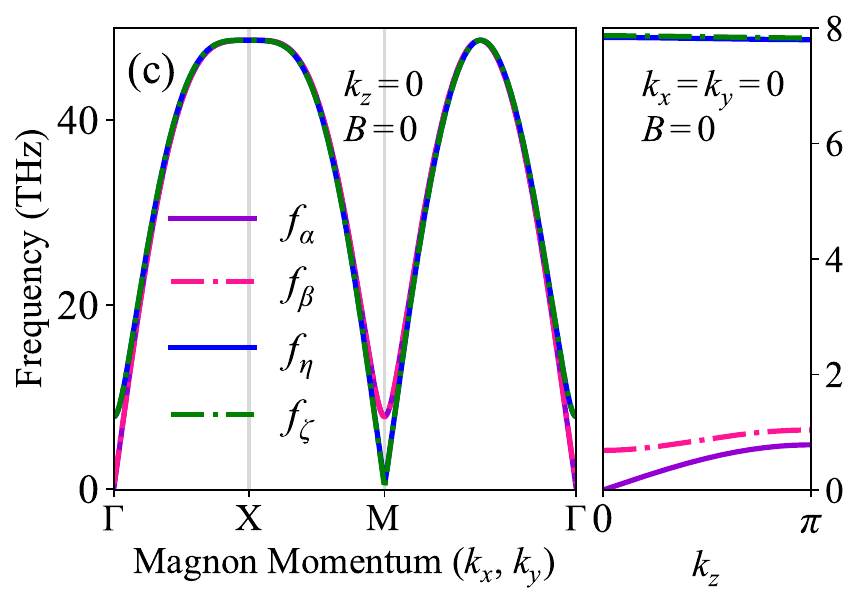}\\
    \includegraphics[width=0.38\textwidth]{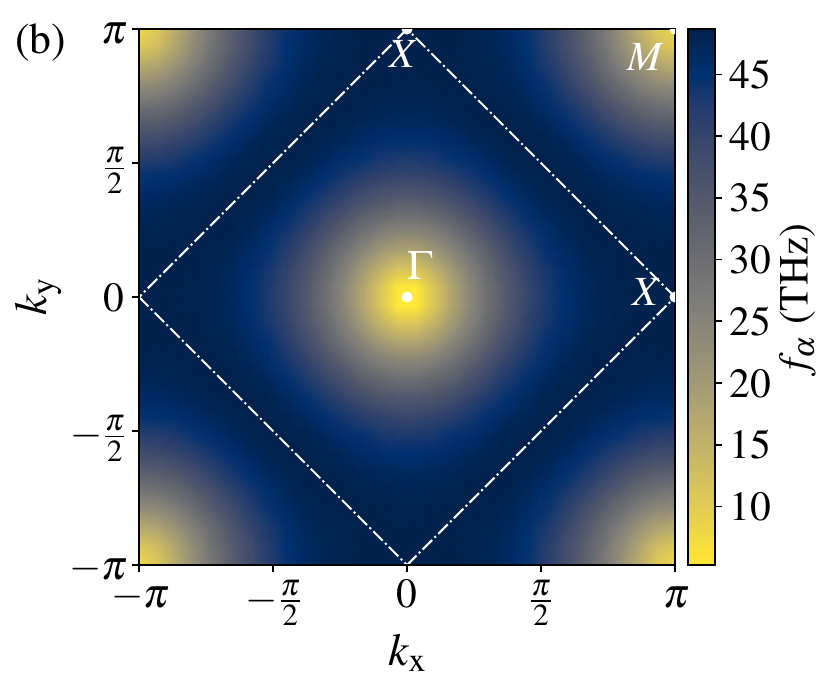}
    \includegraphics[width=0.36\textwidth]{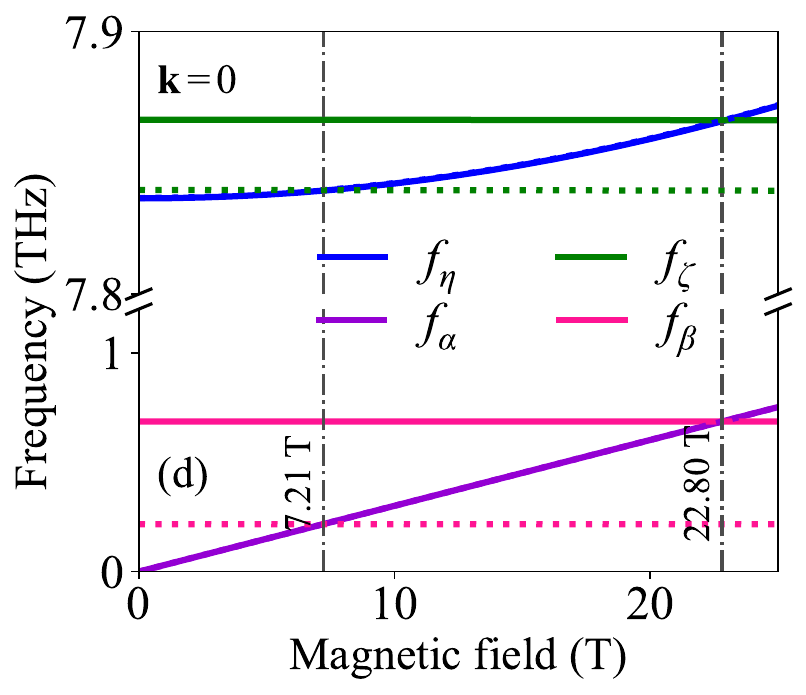}
    \caption{\textbf{Bilayer cuprate crystal structure and its magnon dispersions.} (a) Crystal and spin structure of YBa$_2$Cu$_3$O$_{6+x}$ with the in-plane superexchange $J_\parallel=100\mathrm{meV}$, intra-bilayer exchange $J_{\perp1}=2.6\mathrm{meV}$, and weak inter-bilayer coupling $J_{\perp2}=0.026\mathrm{meV}$ mediated by Cu(1)-O chains. The magnetic field $\mathbf{B}$ induces uniform spin canting within the ab-plane. Structural parameters include the in-plane lattice constant $a$, intra-bilayer Cu-Cu separation $zc$, inter-bilayer spacing $c$, and fractional $z$-coordinate defining the bilayer asymmetry. (b) Computed mode $\alpha$ dispersion across the magnetic Brillouin zone at $k_z = 0$, demonstrating the two-dimensional character with high-symmetry points $\Gamma$, $X$, and $M$ indicated. (c) Complete four-branch magnon spectrum along $\Gamma \to X \to M \to \Gamma$(left) and c-axis dispersion at fixed in-plane momentum (right) in zero field. (d) Magnetic field evolution of $\Gamma$ point frequencies. Solid (dotted) lines are for $\alpha_D = 1.0\times10^{-4}$ ($\alpha_D = 1.0\times10^{-5}$).}  
\label{Fig1}
\end{figure*}

The parent compound YBa$_2$Cu$_3$O$_{6+x}$ (YBCO) exhibits long-range antiferromagnetic order arising from its bilayer CuO$_2$ planes. Each unit cell hosts a CuO$_2$ bilayer separated by yttrium (Fig.~\ref{Fig1}a), with in-plane lattice constant $a\simeq3.86$~\AA, $c$-axis parameter $c\simeq11.7$~\AA, and fractional bilayer offset $z=0.28$, giving a CuO$_2$--CuO$_2$ spacing $zc\simeq3.3$~\AA. Across the oxygenation range that retains long-range AFM order, the N\'eel temperature decreases with $x$. Near $x=0$ $T_N\approx 420$~K, at $x\approx0.1$ $T_N\approx 350\pm50$~K, at $x\approx0.2$ $T_N\approx 250\pm50$~K, and long-range AFM typically vanishes by $x\gtrsim0.35$--$0.40$. The dominant coupling is the AFM superexchange within a plane, $J_{\parallel}\sim(80\text{–}120)$~meV from neutron and Raman scattering~\cite{tranquada1989neutron, abragam2012electron, chen2011angle, vettier1989neutron, chubukov1995resonant}. The bilayer stacking introduces a direct intra-bilayer exchange $J_{\perp1}\approx 2.6$~meV between Cu atoms in adjacent planes, while a much weaker inter-bilayer superexchange $J_{\perp2}\approx 0.026$~meV couples successive bilayers through the Cu--O chain layer. This hierarchy $J_{\parallel}\gg J_{\perp1}\gg J_{\perp2}$ enforces quasi-two-dimensional correlations with limited three-dimensional coherence. Spin–orbit coupling induces a weak easy-plane anisotropy that suppresses out-of-plane fluctuations, confines the ordered moments to the $ab$~plane, and, for definiteness, sets the Néel vector along $\hat{x}$. We apply a static in-plane field along $\hat{\mathbf{y}}$, which induces a transverse canting within the easy plane. The spin-$\tfrac{1}{2}$ Hamiltonian is
\begin{align}
\hat{\mathcal H}_{0} &= \tfrac{1}{2}J_{\parallel} \sum_{\kappa,n,\boldsymbol{a}} 
\Big( \mathbf{S}_{n}^{\kappa}\cdot\mathbf{S}_{n+\boldsymbol{a}}^{\kappa} 
- \alpha_D S_{n}^{z,\kappa} S_{n+\boldsymbol{a}}^{z,\kappa} \Big) \nonumber \\
& + J_{\perp1} \sum_n \mathbf{S}_{n}^{1}\cdot\mathbf{S}_{n}^{2} 
+ J_{\perp2} \sum_n \mathbf{S}_{n}^{1}\cdot\mathbf{S}_{n+\mathbf{c}}^{2}\nonumber\\ 
&+g_{\mathrm{Cu}}\mu_{\mathrm{B}} \sum_{\kappa,n} \mathbf{B}\cdot \mathbf{S}_{n}^{\kappa},
\end{align}
where $\mathbf{S}_{n}^{\kappa}$ denotes the spin operator at site $n$ in layer $\kappa\in\{1,2\}$ and $g_{\mathrm{Cu}}\simeq 2.15$ for Cu$^{2+}$. The transverse field cants the spins by an angle $\chi$, giving the rotated-frame vector $\tilde{\mathbf{S}}$~\cite{vishina2012spin, bostrom2023direct}:
\begin{subequations} 
\label{eq:rotation}
\begin{align}
S_{n,a(b)}^{x} &= \cos\chi\, \tilde{S}_{n,a(b)}^{x} \mp \sin\chi\, \tilde{S}_{n,a(b)}^{y}, \\
S_{n,a(b)}^{y} &= \pm \sin\chi\, \tilde{S}_{n,a(b)}^{x} + \cos\chi\, \tilde{S}_{n,a(b)}^{y}, \\
S_{n,a(b)}^{z} &= \tilde{S}_{n,a(b)}^{z},
\end{align}
\end{subequations}
with the upper (lower) sign corresponding to sublattice $a$ ($b$). In this rotated frame, quantum fluctuations are introduced via the Holstein--Primakoff (HP) transformation. With the N\'eel vector along $\hat{x}$, we quantize on the local $x$-axis and define $\tilde S^{\pm}\equiv \tilde S^{z}\pm i\tilde S^{y}$. On sublattice $a$ (and equivalently $c$) this gives $\tilde S_{n}^{-}=\sqrt{2S}\,a_{n}^{\dagger}$, $\tilde S_{n}^{+}=\sqrt{2S}\,a_{n}$, and $\tilde S_{n}^{x}=S-a_{n}^{\dagger}a_{n}$, while on sublattice $b$ (and equivalently $d$) one has $\tilde S_{n}^{-}=\sqrt{2S}\,b_{n}$, $\tilde S_{n}^{+}=\sqrt{2S}\,b_{n}^{\dagger}$, and $\tilde S_{n}^{x}=-S+b_{n}^{\dagger}b_{n}$. In the bilayer geometry, site $c$ lies above sublattice $b$ but follows the local orientation of $a$, whereas site $d$ lies above $a$ and follows the orientation of $b$. Spin-wave operators are obtained by Fourier transforming to reciprocal space as  $\mathcal{O}_n = N^{-1/2} \sum_{\mathbf{k}} \mathcal{O}_{\mathbf{k}} e^{i s_{\mathcal{O}} \mathbf{k} \cdot \mathbf{R}_{\mathcal{O}n}}$, where $N$ is the number of bilayer unit cells, $\mathcal{O}_{\mathbf{k}}$ denotes the operator on sublattice $\mathcal{O} \in \{a,b,c,d\}$, $\mathbf{R}_{\mathcal{O}n}$ is the site position in cell $n$, and $s_{\mathcal{O}} = +1$ for sublattices $a,c$ and $s_{\mathcal{O}} = -1$ for sublattices $b,d$. As a result, the quadratic spin Hamiltonian in momentum space reads
\begin{align}
\hat{\mathcal H}_{0}&=E_\mathrm{cl}+2J_{\parallel}\sum_{\mathbf{k}}\left\{ \mathcal{J}\left[a_{\mathbf{k}}^{\dagger}a_{\mathbf{k}}+b_{\mathbf{k}}^{\dagger}b_{\mathbf{k}}+c_{\mathbf{k}}^{\dagger}c_{\mathbf{k}}+d_{\mathbf{k}}^{\dagger}d_{\mathbf{k}}\right]\right.\nonumber\\
 &\hspace{4em}+\Delta_{\bf{k}}\left[a_{\mathbf{k}}b_{-\mathbf{k}}^{\dagger}+a_{\mathbf{k}}^{\dagger}b_{-\mathbf{k}}+c_{\mathbf{k}}d_{-\mathbf{k}}^{\dagger}+c_{\mathbf{k}}^{\dagger}d_{-\mathbf{k}}\right]\nonumber\\ 
 &\qquad\qquad+\Pi_{\bf{k}}\left[a_{\mathbf{k}}b_{\mathbf{k}}+a_{\mathbf{k}}^{\dagger}b_{\mathbf{k}}^{\dagger}+c_{\mathbf{k}}d_{\mathbf{k}}+c_{\mathbf{k}}^{\dagger}d_{\mathbf{k}}^{\dagger}\right]\nonumber\\
 &\hspace{4em}+\mathfrak{B}^{2}\left[\gamma_{\perp,{\bf{k}}}\left(a_{\mathbf{k}}d_{-\mathbf{k}}^{\dagger}+b_{\mathbf{k}}^{\dagger}c_{-\mathbf{k}}\right)+\mathrm{H.c.}\right]\nonumber\\ 
 &\hspace{4em}\left.+\left(1-\mathfrak{B}^{2}\right)\left[\gamma_{\perp,{\bf{k}}}\left(a_{\mathbf{k}}d_{\mathbf{k}}+b_{\mathbf{k}}^{\dagger}c_{\mathbf{k}}^{\dagger}\right)+\mathrm{H.c.}\right]\right\},
\end{align}
where $\overline{J}_{\perp} = (J_{\perp1} + J_{\perp2})/2$, $\mathcal{J}=1+\frac{{\overline{J}_{\perp}}}{2J_{\parallel}}$, $\Pi_{\bf{k}}=\left(1-\mathfrak{B}^{2}-\frac{\alpha_{D}}{2}\right)\gamma_{\parallel,{\bf{k}}}$, $\Delta_{\bf{k}}=\left(\mathfrak{B}^{2}-\frac{\alpha_{D}}{2}\right)\gamma_{\parallel,{\bf{k}}}$, and
\begin{align}
\label{eq:gamma_definitions}
\gamma_{\parallel}(\mathbf{k})&=\tfrac{1}{2}\big[\cos(a k_x)+\cos(a k_y)\big], \\
\gamma_{\perp}(\mathbf{k})&=|\gamma_{\perp}(\mathbf{k})|\,e^{i\Phi_{\perp}(\mathbf{k})},
\end{align}
where
\begin{align}
|\gamma_{\perp}(\mathbf{k})| &= \frac{\sqrt{J_{\perp1}^{2}+J_{\perp2}^{2}+2J_{\perp1}J_{\perp2}\cos(ck_{z})}}{4J_{\parallel}}, \nonumber \\  
\Phi_{\perp}(\mathbf{k}) &= \arctan\left[\frac{J_{\perp1}\sin(zck_{z}) - J_{\perp2}\sin((1-z)ck_{z})}{J_{\perp1}\cos(zck_{z}) + J_{\perp2}\cos((1-z)ck_{z})}\right]. \nonumber
\end{align}

The classical ground-state energy per unit cell is $E_{\mathrm{cl}}/N = 2g_{\mathrm{Cu}}\mu_{\mathrm{B}} B\sin\chi - (2J_{\parallel} + \overline{J}_{\perp})(1 - 2\sin^2\chi)$, and minimization with respect to $\chi$ gives the equilibrium canting angle~\cite{maksimov2016field,bostrom2021all,bostrom2023direct}:
\begin{equation}
    \sin\chi = -\frac{g_{\mathrm{Cu}}\mu_{\mathrm{B}} B}{4J_{\parallel}+2\overline{J}_{\perp}} \equiv -\mathfrak{B}.
\end{equation}

In the Nambu basis vector $\left[\psi_{\mathbf{k}}^{\dagger},\psi_{-\mathbf{k}}\right]$ where $\psi_{\mathbf{k}}^{\dagger}=\left[\begin{array}{cccc} a_{\mathbf{k}}^{\dagger} & b_{-\mathbf{k}}^{\dagger} & c_{\mathbf{k}}^{\dagger} & d_{-\mathbf{k}}^{\dagger}\end{array}\right]$, the Hamiltonian adopts the canonical Bogoliubov form:  
\begin{align}
\hat{\mathcal H}_{0}&=J_{\parallel}\sum_{\mathbf{k}}\begin{bmatrix}
\psi_{\mathbf{k}}^{\dagger} & \psi_{-\mathbf{k}}
\end{bmatrix}
\begin{bmatrix}
\mathscr{A}_{\mathbf{k}} & \mathscr{B}_{\mathbf{k}}\\
\mathscr{B}_{\mathbf{k}}^{\dagger} & \mathscr{A}_{-\mathbf{k}}^{*}
\end{bmatrix}
\begin{bmatrix}
\psi_{\mathbf{k}}\\
\psi_{-\mathbf{k}}^{\dagger}
\end{bmatrix}
\end{align}
where the 4$\times$4 matrix blocks $\mathscr{A}_{\mathbf{k}}$ and $\mathscr{B}_{\mathbf{k}}$ encode the complete coupling structure (see Supplementary Information (SI)). Diagonalization proceeds through an eight-dimensional bosonic Bogoliubov transformation that maps the original sublattice operators to normal-mode excitations (\(\alpha_{\pm\mathbf{k}}, \beta_{\pm\mathbf{k}}, \eta_{\pm\mathbf{k}}, \zeta_{\pm\mathbf{k}}\))
\begin{equation}
\begin{bmatrix}
\psi_{\mathbf{k}}\\[2pt]
\psi_{-\mathbf{k}}^{\dagger}
\end{bmatrix}
=U_{\mathbf{k}}
\begin{bmatrix}
\varPsi_{\mathbf{k}}\\[2pt]
\varPsi_{-\mathbf{k}}^{\dagger}
\end{bmatrix},
\qquad\varPsi_{\mathbf{k}}^{\dagger}=\big[\alpha_{\mathbf{k}}^{\dagger},\ \beta_{-\mathbf{k}}^{\dagger},\ \eta_{\mathbf{k}}^{\dagger},\ \zeta_{-\mathbf{k}}^{\dagger}\big].
\end{equation}

The paraunitary matrix $U_{\mathbf{k}}$ satisfies $U_{\mathbf{k}}^{\dagger}\,\chi\,U_{\mathbf{k}}=\chi$, where the Nambu metric $\chi=\sigma^{z}\!\otimes\! I_{4}=\mathrm{diag}(I_{4},-I_{4})$ preserves bosonic commutation relations. This transformation block-diagonalizes the dynamical matrix according to:
\begin{gather}
U_{\mathbf k}^{-1}(\chi H_{\mathbf k})U_{\mathbf k} = \mathrm{diag}\!\big(\Omega_{\mathbf k},-\Omega_{\mathbf k}\big),
\end{gather}
with $\Omega_{\mathbf k}=\mathrm{diag}(E_\alpha,E_\beta,E_\eta,E_\zeta)$. The spectrum exhibits the particle–hole symmetry of bosonic Bogoliubov quasiparticles, yielding four positive-energy magnon branches $E_{\alpha}(\mathbf{k})$, $E_{\beta}(\mathbf{k})$, $E_{\eta}(\mathbf{k})$, and $E_{\zeta}(\mathbf{k})$ that represent the complete set of collective magnetic excitations in the bilayer AFM. The diagonalized Hamiltonian reads:
\begin{equation}
\hat{\mathcal{H}}_{0}=\sum_{\mathbf{k}}\sum_{\tau=\alpha,\beta,\eta,\zeta} E_{\tau}(\mathbf{k})\,\tau_{\mathbf{k}}^{\dagger}\tau_{\mathbf{k}}, 
\label{eq:diagH}
\end{equation}
where the four eigenmode branches correspond to in-plane acoustic ($\alpha$) and optical ($\beta$) modes together with out-of-plane acoustic ($\eta$) and optical ($\zeta$) modes. The magnon dispersions are:
\begin{widetext}
\begin{subequations}
\begin{align}
    E_{\alpha}(\mathbf{k})&=2J_{\parallel}\sqrt{2\mathfrak{B}^{2}|\gamma_{\perp}(\mathbf{k})|\left(\Delta_{\mathbf{k}}+\mathcal{J}+\Pi_{\mathbf{k}}+|\gamma_{\perp}(\mathbf{k})|\right)+\left(\Delta_{\mathbf{k}}+\mathcal{J}\right)^{2}-\left(\Pi_{\mathbf{k}}+|\gamma_{\perp}(\mathbf{k})|\right)^{2}}\qquad\qquad \text{(IPA)},\\
    E_{\beta}(\mathbf{k})&=2J_{\parallel}\sqrt{2\mathfrak{B}^{2}|\gamma_{\perp}(\mathbf{k})|\left(\Delta_{\mathbf{k}}-\mathcal{J}+\Pi_{\mathbf{k}}+|\gamma_{\perp}(\mathbf{k})|\right)+\left(\Delta_{\mathbf{k}}-\mathcal{J}\right)^{2}-\left(\Pi_{\mathbf{k}}+|\gamma_{\perp}(\mathbf{k})|\right)^{2}}\qquad\qquad \text{(IPO)},\\
    E_{\eta}(\mathbf{k})&=2J_{\parallel}\sqrt{-2\mathfrak{B}^{2}|\gamma_{\perp}(\mathbf{k})|\left(\Delta_{\mathbf{k}}+\mathcal{J}+\Pi_{\mathbf{k}}-|\gamma_{\perp}(\mathbf{k})|\right)+\left(\Delta_{\mathbf{k}}+\mathcal{J}\right)^{2}-\left(\Pi_{\mathbf{k}}-|\gamma_{\perp}(\mathbf{k})|\right)^{2}}\qquad\quad \text{(OPA)},\\
    E_{\zeta}(\mathbf{k})&=2J_{\parallel}\sqrt{-2\mathfrak{B}^{2}|\gamma_{\perp}(\mathbf{k})|\left(\Delta_{\mathbf{k}}-\mathcal{J}+\Pi_{\mathbf{k}}-|\gamma_{\perp}(\mathbf{k})|\right)+\left(\Delta_{\mathbf{k}}-\mathcal{J}\right)^{2}-\left(\Pi_{\mathbf{k}}-|\gamma_{\perp}(\mathbf{k})|\right)^{2}} \qquad\quad \text{(OPO)}.
    \label{dispersion}
\end{align}
\end{subequations}
\end{widetext}

The corresponding eigenmodes are Bogoliubov quasiparticles, expressed as linear combinations of sublattice bosons. Explicitly, 
\begin{subequations}
\begin{align}
     \alpha_{\mathbf{k}}&=u_{\alpha\mathbf{k}}^{*}\left(a_{\mathbf{k}}+b_{-\mathbf{k}}\right)+u_{\alpha\mathbf{k}}\left(c_{\mathbf{k}}+d_{-\mathbf{k}}\right)\nonumber\\
     &-v_{\alpha\mathbf{k}}^{*}\left(a_{-\mathbf{k}}^{\dagger}+b_{\mathbf{k}}^{\dagger}\right)-v_{\alpha\mathbf{k}}\left(c_{-\mathbf{k}}^{\dagger}+d_{\mathbf{k}}^{\dagger}\right),\\
     \beta_{\mathbf{k}}&=u_{\beta\mathbf{k}}\left(b_{\mathbf{k}}-a_{-\mathbf{k}}\right)+u_{\beta\mathbf{k}}^{*}\left(d_{\mathbf{k}}-c_{-\mathbf{k}}\right)\nonumber \\
     &-v_{\beta\mathbf{k}}\left(a_{\mathbf{k}}^{\dagger}-b_{-\mathbf{k}}^{\dagger}\right)-v_{\beta\mathbf{k}}^{*}\left(c_{\mathbf{k}}^{\dagger}-d_{-\mathbf{k}}^{\dagger}\right), \\
     \eta_{\mathbf{k}}&=u_{\eta\mathbf{k}}^{*}\left(a_{\mathbf{k}}+b_{-\mathbf{k}}\right)-u_{\eta\mathbf{k}}\left(c_{\mathbf{k}}+d_{-\mathbf{k}}\right)\nonumber \\
     &-v_{\eta\mathbf{k}}^{*}\left(a_{-\mathbf{k}}^{\dagger}+b_{\mathbf{k}}^{\dagger}\right)+v_{\eta\mathbf{k}}\left(c_{-\mathbf{k}}^{\dagger}+d_{\mathbf{k}}^{\dagger}\right), \\
     \zeta_{\mathbf{k}}&=u_{\zeta\mathbf{k}}\left(b_{\mathbf{k}}-a_{-\mathbf{k}}\right)+u_{\zeta\mathbf{k}}^{*}\left(c_{-\mathbf{k}}-d_{\mathbf{k}}\right)\nonumber\\
     &-v_{\zeta\mathbf{k}}\left(a_{\mathbf{k}}^{\dagger}-b_{-\mathbf{k}}^{\dagger}\right)+v_{\zeta\mathbf{k}}^{*}\left(c_{\mathbf{k}}^{\dagger}-d_{-\mathbf{k}}^{\dagger}\right),
\end{align}
\end{subequations}
where the Bogoliubov coefficients for modes $\varrho\in\left\{ \alpha,\eta\right\}$ are
\begin{align*}
    u_{\varrho\mathbf{k}}&=\sqrt{\frac{\mathcal{J}+\Delta_{\mathbf{k}}\pm\mathfrak{B}^{2}|\gamma_{\perp\mathbf{k}}|+E_{\varrho}(\mathbf{k})/2J_{\parallel}}{8E_{\varrho}(\mathbf{k})/2J_{\parallel}}}e^{-i\Phi_{\perp\mathbf{k}}/2},\\
    v_{\varrho\mathbf{k}}&=\sqrt{\frac{\mathcal{J}+\Delta_{\mathbf{k}}\pm\mathfrak{B}^{2}|\gamma_{\perp\mathbf{k}}|-E_{\varrho}(\mathbf{k})/2J_{\parallel}}{8E_{\varrho}(\mathbf{k})/2J_{\parallel}}}e^{-i\Phi_{\perp\mathbf{k}}/2},  
\end{align*}
where the upper (lower) sign corresponds to $\alpha$($\eta$). For modes $\varrho\in\left\{ \beta,\zeta\right\}$ 
\begin{align*}
    u_{\varrho\mathbf{k}}&=\sqrt{\frac{\mathcal{J}-\Delta_{\mathbf{k}}\mp\mathfrak{B}^{2}|\gamma_{\perp\mathbf{k}}|+E_{\varrho}(\mathbf{k})/2J_{\parallel}}{8E_{\varrho}(\mathbf{k})/2J_{\parallel}}}e^{-i\Phi_{\perp\mathbf{k}}/2},\\ 
    v_{\varrho\mathbf{k}}&=\sqrt{\frac{\mathcal{J}-\Delta_{\mathbf{k}}\mp\mathfrak{B}^{2}|\gamma_{\perp\mathbf{k}}|-E_{\varrho}(\mathbf{k})/2J_{\parallel}}{8E_{\varrho}(\mathbf{k})/2J_{\parallel}}}e^{-i\Phi_{\perp\mathbf{k}}/2},
\end{align*}
where the upper (lower) sign corresponds to $\beta$ ($\zeta$). When \(\Phi_{\perp}(\mathbf{k})=0\) (e.g., for \(k_{z}=0\)), all coefficients are real.

Fig.~\ref{Fig1}b displays the zero-field frequency map $f_{\alpha}(k_x,k_y)$ at $k_z=0$, reflecting the square-lattice symmetry of the CuO$_2$ planes. Fig.~\ref{Fig1}c shows the zero-field dispersions of magnon branches along the high-symmetry path $\Gamma\!\to\!X\!\to\!M\!\to\!\Gamma$. As $\gamma_{\parallel}(\mathbf{k})$ evolves from $+1$ at $\Gamma$ to $-1$ at $M$, the branch ordering reverses such that $E_{\alpha}(\Gamma)=E_{\zeta}(M)$ and $E_{\beta}(\Gamma)=E_{\eta}(M)$. Along the magnetic Brillouin-zone boundary—e.g., at $(ak_x,ak_y)=(\pi/2,\pi/2)$ where $\gamma_{\parallel}(\mathbf{k})=0$—the in-plane mixing vanishes and all four branches collapse to a single energy, marking the disappearance of bilayer splitting. The zone-center eigenenergies are 
\begin{subequations} 
\begin{align*}
E_{\alpha}(\Gamma)
&= B_\mathrm{Z}\,\sqrt{1-\frac{\alpha_{D}}{2\mathcal{J}}},\\[3pt]
E_{\beta}(\Gamma)&= 2J_{\parallel}\,\sqrt{\,2\alpha_{D}\,\mathcal{J}
-\frac{\alpha_{D}}{8\mathcal{J}}\frac{B_\mathrm{Z}^{2}}{J_{\parallel}^{2}}\,},\\[3pt]
E_{\eta}(\Gamma)&= \sqrt{\,(2-\alpha_{D})\!\left[\,4J_{\parallel}\overline{J}_{\perp}
+\frac{B_\mathrm{Z}^{2}}{2\mathcal{J}^{2}}\!\left(1-\frac{\overline{J}_{\perp}}{2J_{\parallel}}\right)\right]},\\[3pt]
E_{\zeta}(\Gamma)&= \sqrt{\,\left(\alpha_{D}+\frac{\overline{J}_{\perp}}{J_{\parallel}}\right)\!\left[\,8J_{\parallel}^{2}+\frac{B_\mathrm{Z}^{2}}{4\mathcal{J}^{2}}\!\left(\frac{\overline{J}_{\perp}}{J_{\parallel}}-2\right)\right]},
\end{align*}
\end{subequations} 
here $B_\mathrm{Z} \equiv g_{\mathrm{Cu}}\mu_{\mathrm{B}}B$, at $B=0$ energies reduce to $0$, $2J_{\parallel}\sqrt{2\alpha_D\mathcal{J}}$, $2J_{\parallel}\sqrt{(2-\alpha_D)\,\overline{J}_{\perp}/J_{\parallel}}$, and $2J_{\parallel}\sqrt{2\!\left(\alpha_D+\overline{J}_{\perp}/J_{\parallel}\right)}$, respectively. The field dependence of the $\Gamma$-point magnon modes is shown in Fig.~\ref{Fig1}d. Most importantly, because $\alpha_D\ll\mathcal{J}$, the $\alpha$ mode is weakly influenced by anisotropy and exhibits nearly perfect linear Zeeman scaling, $E_{\alpha}\simeq B$, while remaining gapless at zero field. This hallmark behavior identifies it as the Goldstone mode associated with spontaneous U(1) symmetry breaking. Mode $\eta$ shows negligible anisotropy dependence and acquires only a weak quadratic correction under applied field. By contrast, the anisotropy-dominated $\beta$ and $\zeta$ modes scale as $\propto\sqrt{\alpha_D}$ and remain essentially field-independent. For reference, the Supplementary Information compares YIG and cuprate $\alpha$-mode frequencies. Despite higher damping in cuprates, the $\alpha$ mode shows a steeper Zeeman slope ($\sim$30 vs.\ 28 GHz/T) and coexists with THz optical branches, enabling broadband (GHz–THz) hybrid cavity architectures beyond YIG-only platforms~\cite{sadat2023enhancing}.

In this work, we exploit the gapless $\alpha$ mode as a quantum gateway, coupling it to microwave cavities to open a new frontier in magnonic quantum electrodynamics. Its entanglement with the higher-frequency $\beta$ mode, rooted in their shared Bogoliubov origin~\cite{yuan2020enhancement, mousolou2020hierarchy}, enables coherent transfer of quantum information between microwave and THz scales, effectively turning the bilayer cuprate into a solid-state frequency converter. For the bulk anisotropy $\alpha_D=1.0\times10^{-4}$~\cite{tranquada1989neutron}, the $\beta$ branch lies at $686~\mathrm{GHz}$, with the $\alpha$–$\beta$ crossing at $B=22.8~\mathrm{T}$. At this degeneracy, magneto-optical probes such as MOKE can address the $\beta$ mode while microwave techniques control the $\alpha$ mode, realizing a microwave–THz quantum transducer spanning nearly three orders of magnitude in frequency. This functionality is tunable: reducing anisotropy via epitaxial strain, oxygen stoichiometry, or targeted substitution~\cite{vishina2012spin, IHARA19971973, burlet1998plane, kochelaev2016spin} lowers $f_\beta$ and shifts the crossing into experimentally accessible fields. For $\alpha_D=1.0\times10^{-5}$, the modes resonate at $217~\mathrm{GHz}$ with $B=7.22~\mathrm{T}$, already compatible with emerging THz quantum technologies. Driving the anisotropy further down brings both branches into the microwave domain, enabling complete quantum protocols with standard cavity instrumentation. This tunability positions bilayer cuprates as a uniquely versatile platform, functioning either as a microwave–THz transducer or as an all-microwave architecture for scalable quantum information processing.

\section{Optomagnonic Coupling}
\label{Optomagnonic}
The quantized magnetic field in the cavity follows standard field quantization~\cite{yuan2017magnon,johansen2018nonlocal,xiao2019magnon}:
\begin{equation}
\hat{\mathbf{B}}_{\mathrm{p}}(\mathbf{r})
= \mathrm{i} \sum_{\lambda,\mathbf{k}}
\sqrt{\frac{\mu_{0}\hbar\omega_{\mathbf{k}}}{2V_{\mathrm{B}}}}\,
\Big[(\hat{\mathbf{k}}\times\mathbf{e}_{\lambda})\,\hat{p}_{\lambda\mathbf{k}}\,e^{\mathrm{i}\mathbf{k}\cdot\mathbf{r}}
+ \text{H.c.}\Big],
\label{eq:B_quant}
\end{equation}
where $\hat{p}_{\lambda\mathbf{k}}^{\dagger}$ ($\hat{p}_{\lambda\mathbf{k}}$) creates (annihilates) a photon in the mode characterized by wave vector $\mathbf{k}$, frequency $\omega_{\mathbf{k}}$, and helicity $\lambda=\pm1$, and $V_{\mathrm{B}}$ is the effective cavity mode volume. The propagation direction is $\hat{\mathbf{k}}=\sin\theta\cos\phi\,\hat{\mathbf{x}}
+\sin\theta\sin\phi\,\hat{\mathbf{y}}
+\cos\theta\,\hat{\mathbf{z}}$ with circular polarization vector $\mathbf{e}_{\lambda}= (e_{x}\hat{\mathbf{x}}+e_{y}\hat{\mathbf{y}}+e_{z}\hat{\mathbf{z}})/\sqrt{2}$, where $e_{x}=-\sin\phi-\mathrm{i}\lambda\cos\theta\cos\phi$,
$e_{y}=\cos\phi-\mathrm{i}\lambda\cos\theta\sin\phi$,
and $e_{z}=\mathrm{i}\lambda\sin\theta$. Restricting to a single cavity mode at resonance frequency $\omega_{c}$ with $(\lambda_{0},\mathbf{k}_{0})$, the cavity Hamiltonian reduces to
\(\hat{\mathcal{H}}_{c}=\hbar\omega_{c}\big(\hat{p}^{\dagger}\hat{p}+1/2\big)\),
where $\hat{p}\equiv\hat{p}_{\lambda_{0}\mathbf{k}_{0}}$. The spins couple to the cavity photon field via the Zeeman interaction
\(\hat{\mathcal{H}}_{\mathrm{I}} = g_{\mathrm{Cu}}\mu_{\mathrm{B}}\sum_{n}\hat{\mathbf{B}}_{\mathrm{p}}(\mathbf{r}_{n})\cdot\hat{\mathbf{S}}_{n}\).
After Holstein–Primakoff and Fourier transformations, this yields (see SI)
\begin{align}
\hat{\mathcal{H}}_{\mathrm{I}}
&= \sum_{\mathbf{k}} g_{0}\Big\{
\hat{p}_{\mathbf{k}}\big[\Lambda_{1}(\hat{a}_{-\mathbf{k}}+\hat{c}_{-\mathbf{k}})
+\Lambda_{2}(\hat{a}_{\mathbf{k}}^{\dagger}+\hat{c}_{\mathbf{k}}^{\dagger}) \nonumber\\
&\hspace{4em}
+\Lambda_{3}(\hat{b}_{-\mathbf{k}}^{\dagger}+\hat{d}_{-\mathbf{k}}^{\dagger})
+\Lambda_{4}(\hat{b}_{\mathbf{k}}+\hat{d}_{\mathbf{k}})\big]
+ \text{H.c.}\Big\},
\label{eq:HI_main}
\end{align}
with polarization–canting coefficients
\begin{subequations}
\begin{align}
    \Lambda_{1,2} &= \tfrac{\lambda_0}{4}\left(e_{z} \mp \mathrm{i}e_{y}\cos\chi \pm \mathrm{i}e_{x}\sin\chi\right),\\
    \Lambda_{3,4} &= \tfrac{\lambda_0}{4}\left(e_{z} \mp \mathrm{i}e_{y}\cos\chi \mp \mathrm{i}e_{x}\sin\chi\right),
\end{align}
\end{subequations}
where the upper (lower) sign corresponds to $\Lambda_{1,3}$ ($\Lambda_{2,4}$). The collective coupling strength is $g_{0}=g_{\mathrm{Cu}}\mu_{\mathrm{B}}\sqrt{\mu_{0}\hbar\omega_{c}\rho\eta_{B}}$ with spin density $\rho=N/V_{\mathrm{s}}$ (for YBCO, $\rho\approx 1.1\times10^{28}\,\mathrm{m}^{-3}$~\cite{tranquada1989neutron}) and filling factor $\eta_{B}=V_{\mathrm{s}}/V_{\mathrm{B}}\leq1$ quantifying the overlap between the magnetic sample and cavity field~\cite{huebl2013high}.

\begin{figure*}[htp]
  \centering
  \includegraphics[width=0.34\textwidth]{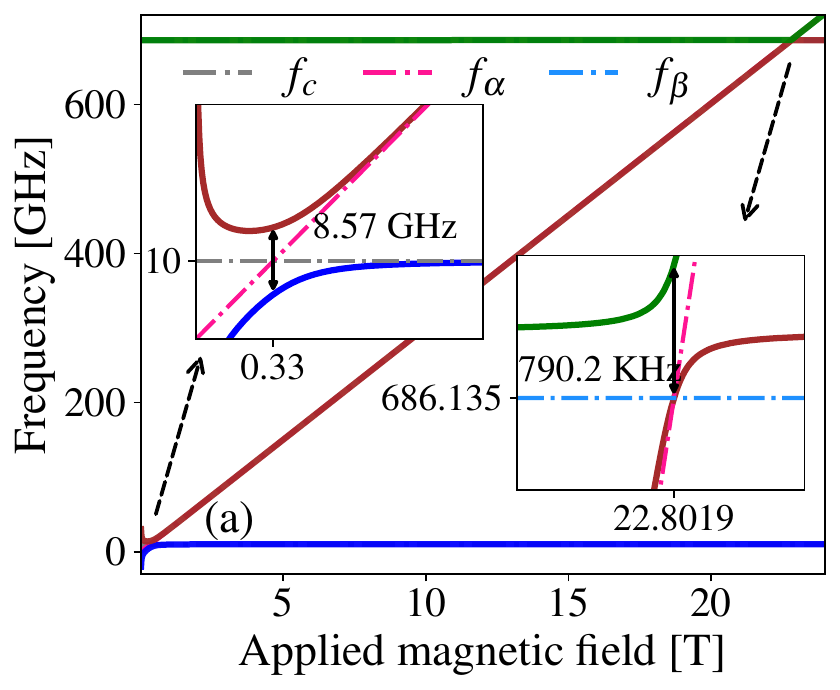}
  \includegraphics[width=0.34\textwidth]{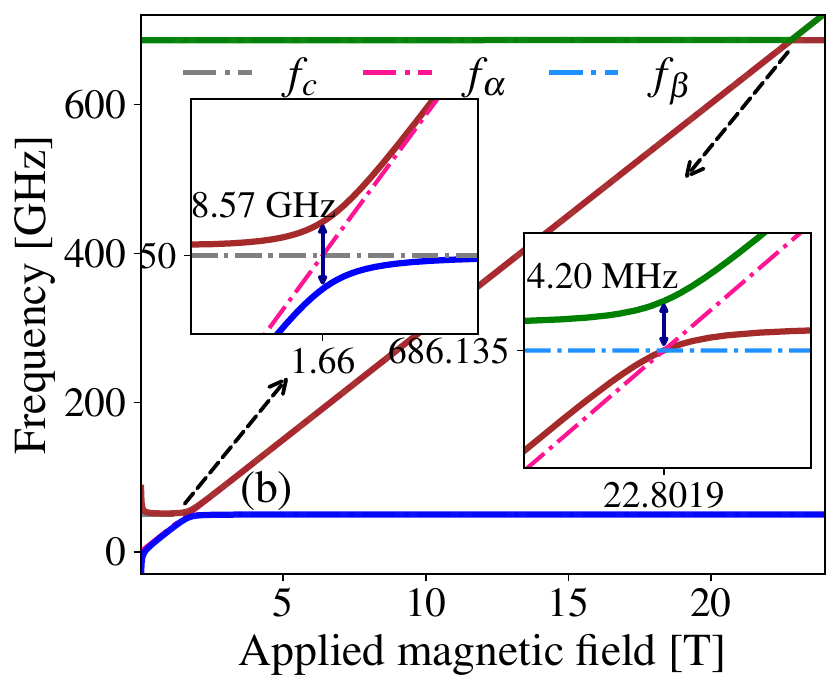}\\
  \includegraphics[width=0.34\textwidth]{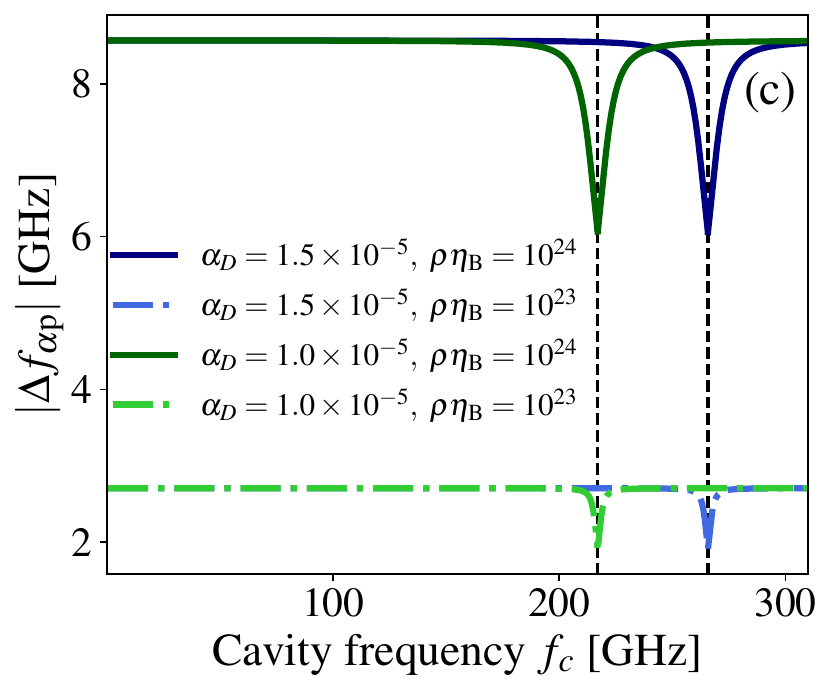}
  \includegraphics[width=0.34\textwidth]{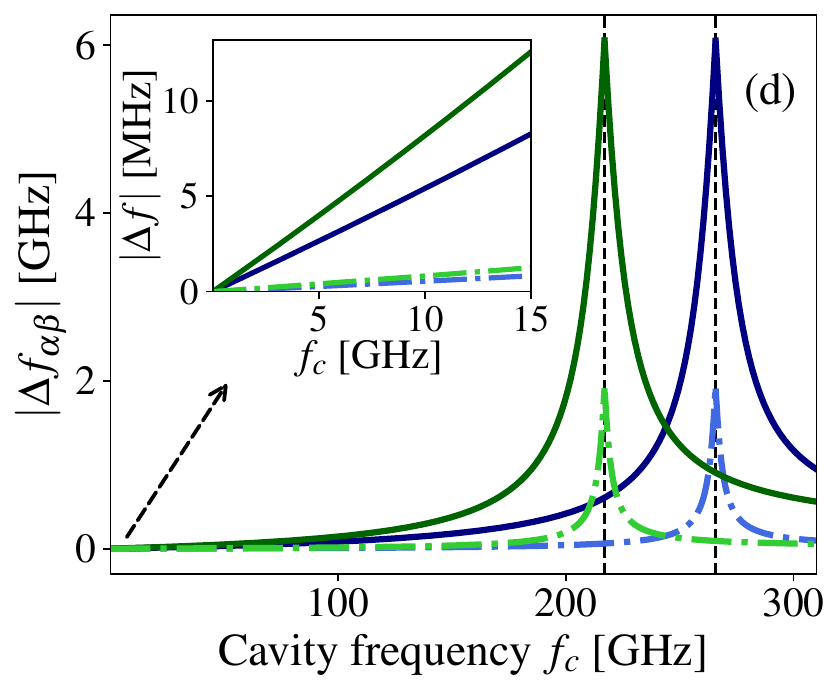}
    \caption{\textbf{Cavity magnon–polariton spectra and gap evolution.} Hybrid eigenfrequencies (solid) of the coupled cavity–AFM system versus magnetic field $B$ for cavity frequencies $f_c=10~\mathrm{GHz}$ (a) and $50~\mathrm{GHz}$ (b), with $\alpha_D=1.0\times10^{-4}$ and $\rho\eta_\mathrm{B}=10^{24}$. Dashed lines show the uncoupled photon ($f_c$) and magnon modes $f_\alpha(B)$ and $f_\beta(B)$. Insets highlight avoided crossings: photon–$\alpha$ splittings of $\Delta f_{\alpha\mathrm{p}}=8.57~\mathrm{GHz}$ in both panels, and $\alpha$–$\beta$ gaps of $\Delta f_{\alpha\beta}=790~\mathrm{kHz}$ (a) and $4.2~\mathrm{MHz}$ (b). (c,d) Magnitudes of the mode splittings, $|\Delta f_{\alpha p}|$ and $|\Delta f_{\alpha\beta}|$, plotted versus $f_c$ for different $\alpha_D$ and $\rho\eta_\mathrm{B}$. The inset in (d) shows the low-$f_c$ regime.}
\label{fig2}
\end{figure*}

After the Bogoliubov transformation at the $\Gamma$ point and neglecting counter-rotating contributions, the photon-magnon interaction reduces to a beam-splitter form,  
\begin{equation}
\hat{\mathcal{H}}_{\mathrm{I}}=\mathcal{G}_{\alpha}\,\hat{\alpha}^{\dagger}\hat{p}+\mathcal{G}_{\beta}\,\hat{\beta}^{\dagger}\hat{p}+\text{H.c.},
\label{eq:H_I_RWA_compact}
\end{equation}
with effective amplitudes  
\begin{align}
\mathcal{G}_{\alpha}&= 2g_{0}\Big[(\Lambda_{2} + \Lambda_{3})\,u_{\alpha}
+(\Lambda_{1} + \Lambda_{4})\,v_{\alpha}\Big],\nonumber\\
\mathcal{G}_{\beta}&= 2g_{0}\Big[(\Lambda_{3} - \Lambda_{2})\,u_{\beta}
+(\Lambda_{1} - \Lambda_{4})\,v_{\beta}\Big],
\label{eq:G_eff_defs}
\end{align}
while $\eta$ and $\zeta$ branches are symmetry-dark at $\Gamma$ for the chosen polarization orientation. Remarkably, the two magnon branches exhibit fundamentally different coupling behaviors: while the $\beta$-mode maintains field-independent coupling, the $\alpha$-mode scales as $|\mathcal{G}_{\alpha}(B)| \propto B^{-1/2}$ (valid away from the strict $B\!\to\!0$ limit). This dual-channel architecture—one fixed, one tunable—is absent in conventional magnets and enables independent control of frequency and coupling strength. Such flexibility allows magnetic optimization of cooperativity, reversible $\alpha$--$\beta$ state transfer, and dynamic multimode entanglement. Most importantly, the system can be continuously tuned between strong- and weak-coupling regimes, positioning bilayer cuprates as a versatile platform for adaptive quantum information processing that bridges microwave and THz domains within a single reconfigurable system.

The linearized Heisenberg–Langevin equations for the annihilation operators are
\begin{subequations}
\begin{align}
\dot{\hat p}&= -\Big(\mathrm{i}\,\omega_c+\tfrac{\kappa_c}{2}\Big)\hat p
               - \mathrm{i}\,g_{\alpha}^{*}\,\hat\alpha
               - \mathrm{i}\,g_{\beta}^{*}\,\hat\beta
               + \sqrt{\kappa_c}\,\hat p_{\mathrm{in}},\\
\dot{\hat\alpha}&= -\Big(\mathrm{i}\,\omega_\alpha+\tfrac{\gamma_\alpha}{2}\Big)\hat\alpha
                   - \mathrm{i}\,g_{\alpha}\,\hat p
                   + \sqrt{\gamma_\alpha}\,\hat\alpha_{\mathrm{in}},\\
\dot{\hat\beta}&=  -\Big(\mathrm{i}\,\omega_\beta+\tfrac{\gamma_\beta}{2}\Big)\hat\beta
                   - \mathrm{i}\,g_{\beta}\,\hat p
                   + \sqrt{\gamma_\beta}\,\hat\beta_{\mathrm{in}},
\end{align}
\end{subequations}
where $g_{\alpha,\beta}\equiv \mathcal G_{\alpha,\beta}/\hbar$ are the photon–magnon coupling rates (rad/s), 
$\kappa_c$, $\gamma_\alpha$, and $\gamma_\beta$ are the total damping rates, and the input operators obey the usual 
$\delta$-correlated noise relations. Seeking solutions $\hat{\mathcal O}(t)\propto e^{-\mathrm{i}\omega t}$ and neglecting inputs leads to the cubic secular equation
\begin{equation}
\omega^{3}+\mathfrak q_{2}\,\omega^{2}+\mathfrak q_{1}\,\omega+\mathfrak q_{0}=0,
\label{roots}
\end{equation}
with $\{\bar\omega_c,\bar\omega_\alpha,\bar\omega_\beta\}
=\{\omega_c,\omega_\alpha,\omega_\beta\}
-\frac{\mathrm i}{2}\{\kappa_c,\gamma_\alpha,\gamma_\beta\}$ and
\begin{subequations}
\begin{align}
\mathfrak q_{2}&= -(\bar\omega_c+\bar\omega_\alpha+\bar\omega_\beta),\\
\mathfrak q_{1}&= \bar\omega_c\bar\omega_\alpha+\bar\omega_c\bar\omega_\beta+\bar\omega_\alpha\bar\omega_\beta
                  -|g_\alpha|^{2}-|g_\beta|^{2},\\
\mathfrak q_{0}&= -\bar{\omega}_{c}\bar{\omega}_{\alpha}\bar{\omega}_{\beta}
                   +\bar{\omega}_{\beta}\,|g_{\alpha}|^{2}
                   +\bar{\omega}_{\alpha}\,|g_{\beta}|^{2}.
\end{align}
\end{subequations}

The three complex roots give the hybrid eigenfrequencies (real parts) and decay rates (imaginary parts). Fig.~\ref{fig2}(a,b) shows the real parts of the hybrid eigenfrequencies versus magnetic field for $f_c=10$ and $50$ GHz. At the photon–$\alpha$ resonance ($f_c=f_\alpha$) the spectrum exhibits an on-resonance vacuum-Rabi splitting of $\Delta f_{\alpha p}=8.57~\mathrm{GHz}$, corresponding to a coupling $g_{\alpha p}=\Delta f_{\alpha p}/2=4.285~\mathrm{GHz}$. With $\kappa_c=0.02~\mathrm{GHz}$ and $\gamma_\alpha=0.20~\mathrm{GHz}$, the cooperativity is $C_{\alpha p}=g_{\alpha p}^{2}/(\kappa_c\gamma_\alpha)\approx 4.59\times10^{3}$, so $g_{\alpha p}>(\kappa_c+\gamma_\alpha)/4$ and the normal modes are well within the strong-coupling (cavity–magnon polariton) regime. At the $\alpha$–$\beta$ degeneracy, the cavity induces a narrow avoided crossing with $g_{\alpha\beta}=\Delta f_{\alpha\beta}/2=0.395~\mathrm{MHz}$ ($f_c=10~\mathrm{GHz}$) and $2.10~\mathrm{MHz}$ ($f_c=50~\mathrm{GHz}$), yielding cooperativities $C_{\alpha\beta}=g_{\alpha\beta}^{2}/(\gamma_\alpha\gamma_\beta)\approx 1.6\times10^{-6}$ and $4.4\times10^{-5}$ (for $\gamma_\alpha=0.2~\mathrm{GHz}$, $\gamma_\beta=0.5~\mathrm{GHz}$), firmly placing the dynamics in the dispersive regime. Here, $|\omega_{\alpha}-\omega_c|,\,|\omega_{\beta}-\omega_c|\gg|g_{\alpha}|,|g_{\beta}|,\kappa_c,\gamma_{\alpha},\gamma_{\beta}$, and a Schrieffer--Wolff elimination of the off-resonant cavity~\cite{xu2019cavity,bravyi2011schrieffer} yields
\begin{equation}
g_{\alpha\beta}^{\mathrm{eff}}
=\tfrac{1}{2}\,g_\alpha g_\beta^{*}
\!\left[\frac{1}{\omega_\alpha-\omega_c}+\frac{1}{\omega_\beta-\omega_c}\right],
\label{eq:geff_general}
\end{equation}
revealing the cavity as a tunable mediator of inter-magnon exchange. Reducing the cavity–magnon detuning $(\omega_m-\omega_c)$—by tuning $f_c$ at the $\alpha$–$\beta$ degeneracy—enhances the cavity-mediated exchange $(\Delta f_{\alpha\beta})$ while $C_{\alpha\beta}\!\ll\!1$ keeps the dynamics dispersive, providing a single practical knob for otherwise negligible inter-magnon coupling, unlike platforms requiring multiple magnets or engineered degeneracies~\cite{li2022coherent,xu2019cavity}.

For a global view of the cavity–frequency response, Figs.~\ref{fig2}(c) and (d) plot the on-resonance gaps $\Delta f_{\alpha p}(f_c)$ and $\Delta f_{\alpha\beta}(f_c)$, where for each $f_c$ the field is set to satisfy $f_c=f_\alpha(B)$ for $\Delta f_{\alpha p}$ and to the $\alpha$–$\beta$ degeneracy for $\Delta f_{\alpha\beta}$. As the cavity approaches the triple resonance condition, $\Delta f_{\alpha p}$ decreases while $\Delta f_{\alpha\beta}$ increases. The two trends meet at the triple resonance $f_c=f_\alpha=f_\beta$, where the system reorganizes into bright and dark superpositions
\[
\hat m=\frac{\mathcal G_\alpha^{*}\hat\alpha+\mathcal G_\beta^{*}\hat\beta}{\mathcal G},\qquad
\hat n=\frac{\mathcal G_\beta\,\hat\alpha-\mathcal G_\alpha\,\hat\beta}{\mathcal G},
\]
where phases are chosen such that $\left\langle m|n\right\rangle =0$ and $\mathcal G=\sqrt{|\mathcal G_\alpha|^2+|\mathcal G_\beta|^2}$ ensures proper normalization. In this basis the Hamiltonian simplifies to
\[
\hat{\mathcal H}_{\text{triple}}
= h f_0\big(\hat p^\dagger\hat p+\hat m^\dagger\hat m+\hat n^\dagger\hat n\big)
+ \,\mathcal G\big(\hat p^\dagger\hat m+\hat m^\dagger\hat p\big),
\]
showing that the cavity couples only to the bright mode $\hat m$ with collective coupling energy $\mathcal G$, while the dark mode $\hat n$ is entirely decoupled. The resulting spectrum consists of two bright polaritons at $f_0\pm\mathcal G/h$ and a dark magnon at $f_0$. Thus, at triple resonance the photon–magnon and magnon–magnon gaps reduce to a single characteristic frequency scale $\mathcal G/h$: the bright–bright separation is $2\,\mathcal G/h$, and the nearest-neighbor gaps are each $\mathcal G/h$, explaining the convergence of the two curves in Fig.~\ref{fig2}(c,d). This three-mode reorganization into bright and dark superpositions is a distinctive feature of bilayer cuprates, beyond the conventional two-mode Rabi doublet of ferrimagnets. The convergence of photon–magnon and magnon–magnon couplings into a single collective scale provides a powerful tuning knob, while the dark magnon—decoupled from cavity dissipation—offers long-lived storage, loss-free coupling, and noise resilience. The bright mode supplies a strong interface for signal conversion, establishing bilayer cuprates as a natural platform for memory and transduction~\cite{gardin2024engineering, zhao2023quantum, falch2025magnon}.

The parameter dependence in Fig.~\ref{fig2}(c,d) highlights two clear tuning knobs. Increasing the magnetic filling factor $\rho\eta_\mathrm{B}$ boosts both $g_{\alpha p}$ and the cavity-mediated $g_{\alpha\beta}$, enhancing the overall coupling strength. By contrast, increasing the anisotropy shifts the triple resonance to higher frequencies and fields. Together, these controls provide a direct way to balance photon–magnon and magnon–magnon interactions in the cavity spectrum. Because $\Gamma$-point AFMR linewidths in cuprates are not reliably established, we adopt a conservative effective linewidth of $\gamma_\alpha=0.2$~GHz and $\gamma_\beta=0.5$~GHz~\cite{wang2024mechanism}.

\section{Spectroscopic Response}
\label{Spectroscopy}
To analyze the transmission spectrum, we introduce a weak probe field at frequency $\omega$ coupled through port~1. In the rotating frame, the drive Hamiltonian takes the form
$i\sqrt{\kappa_1}\!\left(s_{\mathrm{in}}p^\dagger - s_{\mathrm{in}}^{*}p\right)$,
with input-output relations $s_{\mathrm{out},1}=s_{\mathrm{in},1}-\sqrt{\kappa_1}\,p$ and
$s_{\mathrm{out},2}=-\sqrt{\kappa_2}\,p$.
Solving the steady-state equations yields
\begin{equation}
S_{21}(\omega)= -\,\frac{\sqrt{\kappa_1\kappa_2}}{\tfrac{\kappa_c}{2}-i\Delta_c+\Sigma(\omega)},\;
\Sigma(\omega)=\sum_{j\in\{\alpha,\beta\}}\frac{|g_j|^{2}}{\tfrac{\gamma_j}{2}-i\Delta_j},
\end{equation}
where $\kappa_c=\kappa_1+\kappa_2+\kappa_{\mathrm{int}}$ is the total cavity linewidth, $\Delta_c=\omega-\omega_c$ is the cavity detuning, and $\Delta_j=\omega-\omega_j$.
\begin{figure}[t!]
    \centering
    \includegraphics[width=0.47\textwidth]{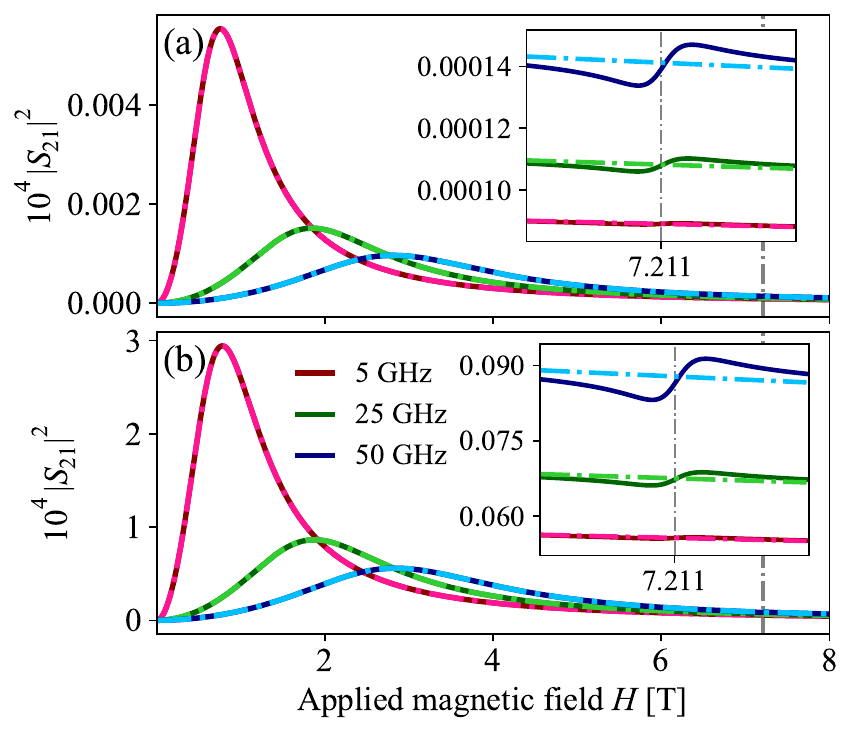}
   \caption{\textbf{Transmission spectra as a function of magnetic field, with the probe frequency locked to the $\alpha$ mode, $f_\mathrm{probe}=f_\alpha(B)$.} (a) Good-cavity regime with $\kappa_{1,2}=0.02$, $\kappa_\mathrm{int}=0.002$. (b) Bad-cavity regime with $\kappa_{1,2}=0.2$, $\kappa_\mathrm{int}=2$. In both magnon linewidths are $\gamma_\alpha=0.2$~GHz and $\gamma_\beta=0.5$GHz and $\alpha_D=1\times10^{-5}$ and $\rho\eta_{\mathrm{B}}=10^{24}$. Solid curves include both bright magnon modes ($\alpha$ and $\beta$), while dashed curves show the case with the $\beta$ mode switched off ($\mathcal{G}_\beta=0$). Curves are color-coded by cavity frequency. Insets enlarge the vicinity of the reference field $B=7.21$~T, where $f_\alpha=f_\beta=f_\mathrm{probe}$, highlighting the emergence of a magnetically tunable Fano-like resonance, disappearing for $\mathcal G_\beta\!\to\!0$.}
    \label{Fig3}
\end{figure}

\begin{figure*}[t]
    \centering
    \includegraphics[width=\linewidth]{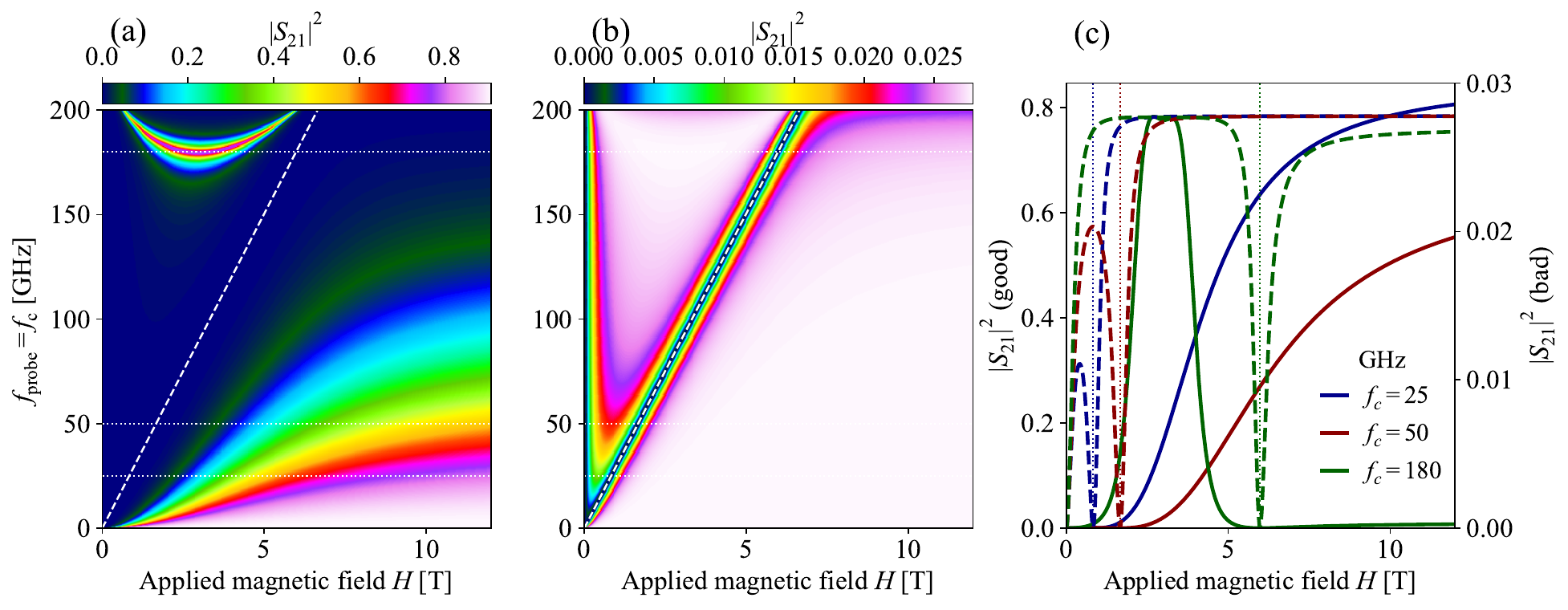}
    \caption{\textbf{Field-tuned transmission with the probe locked to the cavity ($f_{\mathrm{probe}}=f_c$).} (a) Good-cavity ($\kappa_c\!\ll\!\gamma_\alpha$) and (b) bad-cavity ($\kappa_c\!\gg\!\gamma_\alpha$) maps of $|S_{21}|^{2}$ versus $H$ and $f_c$. Dashed line trace $f_{\alpha}(H)$. (c) Line cuts at selected $f_c$; vertical guides mark fields $H$ where $f_c=f_{\alpha}(H)$. Parameters are as in Fig.~\ref{Fig3}.}
    \label{Fig4}
\end{figure*}

Fig.~\ref{Fig3} shows the transmission versus magnetic field with the probe locked to the \(\alpha\) mode, \(f_{\mathrm{probe}}=f_{\alpha}(B)\), for fixed cavity frequencies \(f_{c}\). In this configuration, the \(\alpha\) branch contributes a purely real (dissipative) self-energy, $\Sigma_{\alpha}(B)=\frac{|g_{\alpha}(B)|^{2}}{\gamma_{\alpha}/2}$, which typically satisfies \(\Sigma_{\alpha}>\kappa_{c}\) and therefore sets the effective linewidth. The \(\beta\) branch, pinned near \(f_{\beta}\simeq 216.98~\mathrm{GHz}\), is far detuned: \(\Re[\Sigma_{\beta}]\) adds a small dissipative load and \(\Im[\Sigma_{\beta}]\) a dispersive shift; numerical control study setting \(g_{\beta}\!\to\!0\) confirms that its influence is negligible at low fields (i.e., away from the \(\alpha\)–\(\beta\) degeneracy). Since $|g_{\alpha}| \propto \sqrt{f_{c}/B}$, higher cavity frequencies strengthen $\Sigma_{\alpha}$, broadening the linewidth and reducing the transmission amplitude. This cavity-frequency control of dissipation enables broadband magnon spectroscopy, tunable linewidths, and reconfigurable magnon–photon interfaces. Finite $\mathcal{G}_\beta$ produces a clear Fano-type interference at the $\alpha$–$\beta$ degeneracy (where $f_{\mathrm{probe}}=f_\alpha=f_\beta$ and, in our model, $\mathcal{G}_\alpha\!\approx\!\mathcal{G}_\beta$). As $f_c$ increases, the interference visibility grows while the field width changes only weakly, consistent with the Schrieffer–Wolff (Eq.\ref{eq:geff_general}). Direct microwave access to the $\beta$ mode near $f_\beta\simeq 216.98~\mathrm{GHz}$ is impractical; this can be addressed by (i) lowering $\alpha_D$ to shift the crossing below $100~\mathrm{GHz}$, or (ii) exploiting coherent $\alpha$–$\beta$ coupling for state transfer—mapping states from the field-tunable $\alpha$ mode onto the robust $\beta$ mode for optical retrieval. Increasing $f_c$ thus strengthens $\alpha$–$\beta$ hybridization, a practical knob for reconfigurable filtering and microwave–optical transduction (with ultimate efficiency set by the engineered optical interface)~\cite{liu2025integrated, gollwitzer2021connecting}. In the bad-cavity regime [Fig.~\ref{Fig3}(b)], the qualitative response is preserved, with only the transmission level enhanced, while the linewidth remains unchanged as it is set by $\Sigma_{\alpha}$.

Fig.~\ref{Fig4} shows the transmission with the probe locked to the cavity, $f_{\mathrm{probe}}=f_c$, as a function of magnetic field $B$. At low $B$ the $\alpha$–cavity loading is strong ($\Re\Sigma_\alpha\!\propto\!|\mathcal G_\alpha|^2\!\sim\!1/B$), and $|S_{21}|^2$ is strongly suppressed well below the bare baseline. As $B$ increases, $|\mathcal G_\alpha|$ weakens and the detuning $|f_c-f_\alpha(B)|$ grows; the loading collapses and the signal rises toward the bare-cavity value $4\kappa_1\kappa_2/\kappa_c^2$. In the bad-cavity limit, field-tunable notch appears at $f_\alpha(B)=f_c$. Increasing \(f_{c}\) both strengthens the \(\alpha\)-induced feature \((|\mathcal{G}_{\alpha}|\!\propto\!\sqrt{f_{c}})\) and shifts the saturation to higher \(B\) in both regimes; as \(f_{c}\!\to\! f_{\beta}\), weak \(\beta\)-mode dressing introduces a small dispersive background. Together, these trends enable a compact, field-programmable microwave notch/transparency element, providing a practical spectroscopic handle on the \(\alpha\)-mode dispersion, damping, and magnetic filling—without requiring a scan of the probe frequency.

\begin{figure}[h!]
    \centering
    \includegraphics[width=0.49\textwidth]{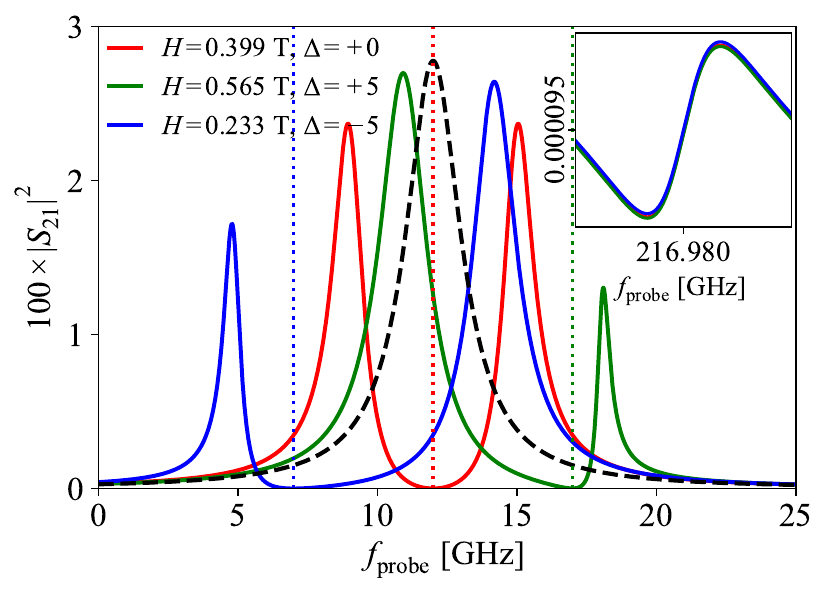}
    \caption{\textbf{Transmission vs probe frequency at fixed cavity frequency $f_c=12.0$ GHz for magnon–cavity detunings $\Delta_{\alpha p}=0,\pm5$ GHz.} Solid curves denote the coupled system, dashed curves the bare cavity, and dotted lines locate $f_\alpha(B)$. Parameters match the bad-cavity regime in Fig.~\ref{Fig3}.}
\label{Fig5}
\end{figure}

Fig.~\ref{Fig5} shows the transmission versus probe frequency at fixed $f_c=12~\mathrm{GHz}$ while the field sets $\Delta_{\alpha p}\!\equiv\! f_c-f_\alpha(B)\in\{0,+5,-5\}\,\mathrm{GHz}$. At $\Delta_{\alpha p}=0$ a symmetric normal-mode doublet appears with splitting $2g_{\alpha p}\!\approx\!6.10~\mathrm{GHz}$ (peaks at $8.950$ and $15.047$ GHz); a fit gives $C=4g_{\alpha p}^{2}/(\kappa_c\gamma_\alpha)\approx7.6\times10^{1}$ and satisfies the resolvable-splitting criterion $g_{\alpha p}>\lvert \kappa_c-\gamma_\alpha\rvert/4$, with polariton linewidths $(\kappa_c+\gamma_\alpha)/2\simeq1.3~\mathrm{GHz}$. For $|\Delta_{\alpha p}|>0$ the $\alpha$ mode induces a dispersive pull $f_{\mathrm{eff}}\simeq f_c+\mathrm{Im}\,\Sigma_\alpha(f_c)$, shifting the cavity-like peak up (down) for $\Delta_{\alpha p}>0$ $(<0)$ and producing a Fano-like feature near $f_\alpha$ whose asymmetry reverses with the sign of $\Delta_{\alpha p}$. A faint, ultranarrow line at $f_\beta$ arises from a virtual-photon pathway via the off-resonant cavity, with power $\propto |g_\beta|^2/\big[(f_\beta-f_c)^2+(\kappa_c/2)^2\big]$. Together, these effects yield a field-programmable microwave element (band-splitter, notch, or dispersive shifter) and enable single-scan extraction of $g_{\alpha p}$, $\kappa_c$, and $\gamma_\alpha$, demonstrating GHz-readout of a sub-THz magnon without retuning; in the good-cavity limit the polariton peaks narrow markedly, supporting high-$Q$ band splitting, filtering, and precision spectroscopy~\cite{wang2018magnon, boventer2020control, noh2023strong, stiewe2022spintronic}.

\section{Conclusion}
\label{sec:conclusion}

Bilayer cuprate antiferromagnets—exemplified by YBa$_2$Cu$_3$O$_{6+x}$—offer a reconfigurable cavity–magnon platform with unusually independent control of frequency and coupling. From a bilayer spin model constrained by neutron data, we identify four magnon branches and focus on two most relevant for cavity physics: a gapless, Zeeman-linear $\alpha$ (Goldstone) mode and an optical $\beta$ mode with $f_\beta\!\propto\!\sqrt{\alpha_D}$, allowing GHz–to–THz placement via anisotropy. Projecting the Zeeman interaction onto a single cavity mode yields two complementary channels: a tunable $|\mathcal G_\alpha|\!\propto\!\sqrt{f_c/B}$ pathway and a nearly field-independent $\beta$ pathway. This duality produces strong photon–$\alpha$ coupling on resonance, a narrow cavity-mediated $\alpha$–$\beta$ exchange in the dispersive regime, and, at triple resonance, a bright/dark reorganization characterized by a single collective scale $\mathcal G$.

The computed eigenfrequencies show a clear vacuum–Rabi splitting at $f_c=f_\alpha$, evidencing coherent photon–$\alpha$ coupling, and a dispersive, cavity-mediated $\alpha$–$\beta$ avoided crossing at their degeneracy. We theoretically analyze the cavity transmission in three configurations. (i) With the probe locked to $\alpha$, the cavity acquires a purely dissipative self-energy, enabling magnetic and cavity-frequency control of linewidth and transmission; near the $\alpha$–$\beta$ crossing, a Fano-like interference appears. (ii) With the probe locked to the cavity, field-tunable notches (bad-cavity limit) or smooth recovery (good-cavity limit) emerge from $\alpha$ loading. (iii) Frequency sweeps at fixed $f_c$ with controlled detuning show a symmetric Rabi doublet at zero detuning and, away from resonance, a dispersive cavity pull with a sign-reversing Fano-like feature near $f_\alpha$; a faint, ultranarrow line at $f_\beta$ persists via a virtual-photon pathway. The predicted splittings, dispersive shifts, and interference lineshapes provide concrete experimental benchmarks and—contingent on damping and filling—enable programmable filtering and—given suitable interfaces—routes toward state storage and microwave-to-THz transduction.

\bibliography{main}
\clearpage
\end{document}